\documentclass[12pt,a4paper]{article}
\usepackage{amsmath,amssymb,graphicx,epsfig,cite,color}
\usepackage[left=2.7cm,right=2.7cm,top=2.5cm,bottom=2.5cm]{geometry}
\usepackage{subfigure}
\usepackage{multirow}
\usepackage[table]{xcolor}
\usepackage{booktabs}

\usepackage{xcolor,colortbl}

\usepackage{graphicx}
\usepackage{lscape}
\usepackage{appendix}
\usepackage{multirow}
\usepackage[colorlinks=true,linkcolor=blue,citecolor=blue,urlcolor=blue]{hyperref}%
\allowdisplaybreaks[1]

\begin{document}

	\renewcommand{\textfraction}{0.2}    
	\renewcommand{\topfraction}{0.8}   
	
	\renewcommand{\bottomfraction}{0.4}   
	\renewcommand{\floatpagefraction}{0.8}
	\newcommand\mysection{\setcounter{equation}{0}\section}
	
\def\OEE{\Omega_{\rm IB}}

\newcommand{\nnl}{\nonumber\\}
\newcommand{\kk}[1]{_{(#1)}}
\newcommand{\kkx}[2]{_{#1(#2)}}
\newcommand{\parity}{\mathbf{P}}
\newcommand{\mcD}{\mathcal{D}}
\newcommand{\mcE}{\mathcal{E}}
\newcommand{\mcG}{\mathcal{G}}
\newcommand{\mcL}{\mathcal{L}}
\newcommand{\mcQ}{\mathcal{Q}}
\newcommand{\mcU}{\mathcal{U}}
\newcommand{\fslash}[1]{#1\!\!\!/}
\newcommand{\nor}{\frac{n}{R}}
\newcommand{\norsq}{\frac{n^2}{R^2}}
\newcommand{\GeV}{~\text{GeV}}
\newcommand{\dpkt}{\;:\quad}
\newcommand{\sw}{s_w}
\newcommand{\cw}{c_w}
\newcommand{\RE}{{\rm Re}}
\newcommand{\IM}{{\rm Im}}
\newcommand{\vcb}{|V_{cb}|}
\newcommand{\vtd}{|V_{td}|}
\newcommand{\vub}{|V_{ub}/V_{cb}|}
\newcommand{\vts}{|V_{ts}|}
\newcommand{\vus}{|V_{us}|}
\newcommand{\tvs}{\vbox{\vskip 3mm}}
\newcommand{\svs}{\vbox{\vskip 5mm}}
\newcommand{\mvs}{\vbox{\vskip 8mm}}
\newcommand{\msvs}{\vbox{\vskip 7mm}}
\def\ss{\langle \bar s s \rangle}
\def\qq{\langle \bar q q \rangle}
\def\mixedss{\langle \bar s \sigma g_s G s \rangle}
\newcommand{\be}{\begin{equation}}
\newcommand{\ee}{\end{equation}}
\newcommand{\bea}{\begin{eqnarray}}
\newcommand{\eea}{\end{eqnarray}}
\newcommand{\nn}{\nonumber}
\newcommand{\dd}{\displaystyle}
\newcommand{\bra}[1]{\left\langle #1 \right|}
\newcommand{\ket}[1]{\left| #1 \right\rangle}
\newcommand{\spur}[1]{\not\! #1 \,}
\def\R1{\varepsilon_1}
\def\E8{\varepsilon_8}
\def\gat{\tilde{\gamma}}
\def\gh{\hat{g}}
\def\gt{\tilde{g}}
\def\gah{\hat{\gamma}}
\def\ga{\gamma}
\def\gaf{\gamma_{5}}
\def\lb{\Lambda_b}
\def\ll{\Lambda}
\def\mb{m_{\Lambda_b}}
\def\ml{m_\Lambda}
\def\s1{\hat s}
\def\ds{\displaystyle}
\def\eps{\varepsilon}
\def\epe{\varepsilon'/\varepsilon}
\def\as{\alpha_s}
\newcommand{\eqn}{\ref}
\def\Heff{{\cal H}_{\rm eff}}
\newcommand{\mt}{m_{\rm t}}
\newcommand{\mtb}{\overline{m}_{\rm t}}
\newcommand{\mcb}{\overline{m}_{\rm c}}
\newcommand{\mc}{m_{\rm c}}
\newcommand{\ms}{m_{\rm s}}
\newcommand{\md}{m_{\rm d}}
\newcommand{\mn}{m_{\rm n}}
\newcommand{\mw}{M_{\rm W}}
\newcommand{\mz}{M_{\rm Z}}

\newcommand{\gev}{\, {\rm GeV}}
\newcommand{\mev}{\, {\rm MeV}}
\newcommand{\bsi}{B_6^{(1/2)}}
\newcommand{\bei}{B_8^{(3/2)}}
\newcommand{\Lms}{\Lambda_{\overline{\rm MS}}}
\newcommand{\bsg}{$b \to s \gamma$ }
\newcommand{\Bsg}{$B \to X_s \gamma$ }
\newcommand{\newsection}[1]{\section{#1}\setcounter{equation}{0}}
\newcommand{\bd}{\begin{displaymath}}
\newcommand{\ed}{\end{displaymath}}
\newcommand{\aem}{\alpha}
\newcommand{\Bsee}{$B \to X_s e^+ e^-$ }
\newcommand{\bsee}{$b \to s e^+ e^-$ }
\newcommand{\bcenu}{$b \to c e \bar\nu $ }

\newcommand{\ord}{{\cal O}}
\newcommand{\order}{{\cal O}}
\newcommand{\f}{\frac}
\newcommand{\Ctilde}{\tilde{C}}
\newcommand{\kpnn}{K^+\rightarrow\pi^+\nu\bar\nu}
\newcommand{\kpn}{K^+\rightarrow\pi^+\nu\bar\nu}
\newcommand{\klpn}{K_{\rm L}\rightarrow\pi^0\nu\bar\nu}
\newcommand{\klpnn}{K_{\rm L}\rightarrow\pi^0\nu\bar\nu}
\newcommand{\klm}{K_{\rm L} \to \mu^+\mu^-}
\newcommand{\kmm}{K_{\rm L} \to \mu^+ \mu^-}
\newcommand{\kpe}{K_{\rm L} \to \pi^0 e^+ e^-}

\setlength{\unitlength}{1mm}
\textwidth 16.3 true cm
\textheight 23.0 true cm
\topmargin -0.8 true in
\oddsidemargin 0.00 true in
\def\theequation{\thesection.\arabic{equation}}
\def\OEE{\Omega_{\rm IB}}

\def\R1{\varepsilon_1}
\def\E8{\varepsilon_8}
\def\gat{\tilde{\gamma}}
\def\gh{\hat{g}}
\def\gt{\tilde{g}}
\def\gah{\hat{\gamma}}
\def\ga{\gamma}
\def\gaf{\gamma_{5}}
\def\eps{\varepsilon}
\def\epe{\varepsilon'/\varepsilon}
\def\as{\alpha_s}
\def\Heff{{\cal H}_{\rm eff}}

\newcommand{\imlt}{\IM\lambda_t}
\newcommand{\relt}{\RE\lambda_t}
\newcommand{\relc}{\RE\lambda_c}
\renewcommand{\baselinestretch}{1.3}
\renewcommand{\thefootnote}{\fnsymbol{}}
\def\ds{\displaystyle}
\def\beq{\begin{equation}}
\def\eeq{\end{equation}}
\def\bea{\begin{eqnarray}}
\def\eea{\end{eqnarray}}
\def\beeq{\begin{eqnarray}}
\def\eeeq{\end{eqnarray}}
\def\ve{\vert}
\def\vel{\left|}
\def\ver{\right|}
\def\nnb{\nonumber}
\def\ga{\left(}
\def\dr{\right)}
\def\aga{\left\{}
\def\adr{\right\}}
\def\lla{\left<}
\def\rra{\right>}
\def\rar{\rightarrow}
\def\nnb{\nonumber}
\def\la{\langle}
\def\ra{\rangle}
\def\ba{\begin{array}}
	\def\ea{\end{array}}
\def\tr{\mbox{Tr}}
\def\ssp{{\Sigma^{*+}}}
\def\sso{{\Sigma^{*0}}}
\def\ssm{{\Sigma^{*-}}}
\def\xis0{{\Xi^{*0}}}
\def\xism{{\Xi^{*-}}}
\def\qs{\la \bar s s \ra}
\def\qu{\la \bar u u \ra}
\def\qd{\la \bar d d \ra}
\def\qq{\la \bar q q \ra}
\def\gGgG{\la g^2 G^2 \ra}
\def\q{\gamma_5 \not\!q}
\def\x{\gamma_5 \not\!x}
\def\g5{\gamma_5}
\def\sb{S_Q^{cf}}
\def\sd{S_d^{be}}
\def\su{S_u^{ad}}
\def\ss{S_s^{??}}
\def\sbp{{S}_Q^{'cf}}
\def\sdp{{S}_d^{'be}}
\def\sup{{S}_u^{'ad}}
\def\ssp{{S}_s^{'??}}
\def\sig{\sigma_{\mu \nu} \gamma_5 p^\mu q^\nu}
\def\fo{f_0(\frac{s_0}{M^2})}
\def\ffi{f_1(\frac{s_0}{M^2})}
\def\fii{f_2(\frac{s_0}{M^2})}
\def\O{{\cal O}}
\def\sl{{\Sigma^0 \Lambda}}
\def\es{\!\!\! &=& \!\!\!}
\def\ap{\!\!\! &\approx& \!\!\!}
\def\ar{&+& \!\!\!}
\def\ek{&-& \!\!\!}
\def\kek{\!\!\!&-& \!\!\!}
\def\cp{&\times& \!\!\!}
\def\se{\!\!\! &\simeq& \!\!\!}
\def\eqv{&\equiv& \!\!\!}
\def\kpm{&\pm& \!\!\!}
\def\kmp{&\mp& \!\!\!}
\def\simlt{\stackrel{<}{{}_\sim}}
\def\simgt{\stackrel{>}{{}_\sim}}

\newcommand{\me}[1]{\langle#1\rangle}
\newcommand{\al}{\alpha_s}

\newcommand{\kpiee}{$K_L \ra \pi^0 e^+ e^-$ }
\renewcommand{\textfraction}{0}
\renewcommand{\topfraction}{0.95}
\setcounter{topnumber}{2}
\renewcommand{\bottomfraction}{0.95}
\setcounter{bottomnumber}{2}
\renewcommand{\floatpagefraction}{0.95}
\setcounter{totalnumber}{3}

\def\baeq{\begin{appeq}}     \def\eaeq{\end{appeq}}  
\def\baeeq{\begin{appeeq}}   \def\eaeeq{\end{appeeq}}
\newenvironment{appeq}{\beq}{\eeq}   
\newenvironment{appeeq}{\beeq}{\eeeq}
\def\bAPP#1#2{
\markright{APPENDIX #1}
\addcontentsline{toc}{section}{Appendix #1: #2}
\medskip
\medskip
\begin{center}      {\bf\LARGE Appendix #1 :}{\quad\Large\bf #2}
\end{center}
\renewcommand{\thesection}{#1.\arabic{section}}
\setcounter{equation}{0}
\renewcommand{\thehran}{#1.\arabic{hran}}
\renewenvironment{appeq}
{  \renewcommand{\theequation}{#1.\arabic{equation}}
\beq}{\eeq}
\renewenvironment{appeeq}
{  \renewcommand{\theequation}{#1.\arabic{equation}}
\beeq}{\eeeq}
\nopagebreak \noindent}
	
\def\eAPP{\renewcommand{\thehran}{\thesection.\arabic{hran}}}
	
\renewcommand{\theequation}{\arabic{equation}}
\newcounter{hran}
\renewcommand{\thehran}{\thesection.\arabic{hran}}

\title{
		{\Large
			{\bf
				Radiative $\Xi_{b}^{-}\rightarrow \Xi^{-}\gamma$ decay
	}}}

\author{\vspace{1cm}\\
		{\small
			A. R. Olamaei$^{1}$ and K. Azizi$^{2,3,}$	 }\footnote{Corresponding author} \\
		{\small$^1$  Department of Physics, Jahrom University,  P.~ O.~ Box 74137-66171, Jahrom, Iran}\\
		{\small $^2$ Department of Physics, University of Tehran, North Karegar Avenue, Tehran 14395-547, Iran}\\
		{\small $^3$ Department of Physics, Do\v{g}u\c{s} University, Dudullu-\"{U}mraniye, 34775
Istanbul, Turkey}			
	} 
	
\date{}

\begin{titlepage}
\maketitle
\thispagestyle{empty}
	
\begin{abstract}
Recently, the LHCb Collaboration performed first search for the rare radiative  $\Xi_{b}^{-}\rightarrow \Xi^{-}\gamma$ decay  and put  an upper limit, ${\cal B}(\Xi_{b}^{-}\to \Xi^{-}\gamma) < 1.3 \times 10^{-4}$, on its branching ratio. The measurement agrees well with existing theory prediction using SU(3) flavor symmetry method, but shows a slight tension with the previous prediction from light-cone sum rules. Inspired by this, we investigate this decay as well as other radiative decays of  $\Xi_b^{0(-)}(\Xi^{'-}_{b})$ to $\Xi^{0(-)}$ and $\Sigma^{0(-)}$ baryons using the form factors calculated from  light-cone QCD sum rules in full theory. we obtain  $ {\cal B}(\Xi_{b}^{-}\to \Xi^{-}\gamma)=1.08^{+0.63}_{-0.49} \times 10^{-5} $, which lies below the upper limit set by LHCb and is consistent with flavor-symmetry driven prediction. Our predictions on other channels may be checked in experiment and by other phenomenological approaches.
\end{abstract}
	
\end{titlepage}
\section{Introduction}
	
Based on the successful quark model, the ordinary hadrons are composed of either  $q\bar{q}$ (meson) or $qqq/\bar{q}\bar{q}\bar{q}$ (baryon/antibaryon) bound states. Among baryons, the ones consist of one heavy quark ($b$ or $c$) are of much interest. They can play the role of a rich “laboratory” for theoretical studies. 
In the limit of infinite mass for the heavy quark ($ m_{Q}\rightarrow\infty $), one can classify the single heavy baryons due to the total flavor-spin wave function of the two remaining light quarks, which has to be symmetric because their color wave function is anti-symmetric. This leads to two different representations ($\textbf{3} \otimes \textbf{3}=\overline{\textbf{3}} \oplus \textbf{6}$) for the ground state of heavy baryons. 
Hence, they are members of either sextet of flavor symmetric state $ \textbf{6} $ with $J^P = \frac{1}{2}^{+}$/$J^P = \frac{3}{2}^{+}$ for total spin-parity of the ground state, or triplet of flavor anti-symmetric state $  \overline{\textbf{3}} $ with ground state spin-parity of $J^P = \frac{1}{2}^{+}$. 

Several phenomenological methods are exploited to perform extensive theoretical studies on the various properties of spin-1/2 heavy baryons, including chiral perturbation theory \cite{Savage}, quark model \cite{Roberts,Karliner}, heavy quark effective theory (HQET) \cite{Dai,Liu,Korner}, hypercentral approach \cite{Ghalenovi1,Ghalenovi2,Patel1,Patel2}, relativistic (constituent) quark model \cite{Ebert1,Ebert2,Ebert3,Migura,Gerasyuta1,Gerasyuta2,Gerasyuta3,Garcilazo}, quark potential model \cite{Capstick}, Feynman-Hellman theorem \cite{Roncaglia}, lattice QCD simulation \cite{Brown,Mathur,Lewis,Bahtiyar}, symmetry-preserving treatment of a vector$\times$vector contact interaction model \cite{Yin}, chiral quark-soliton model \cite{Kim},  QCD sum rules \cite{Shuryak,Bagan,Azizi,Wang1,Wang2,Wang3,Zhang1,Zhang2,Agaev1,Agaev2,Agaev3,Agaev4,Azizi:2013fra,Azizi:2014uxa},  etc.

Thanks to recent progresses in experiments, almost all of the ground states of single heavy baryons are observed \cite{PDG} and investigations on their excited states are ongoing and remarkably important. Therefore, more studies on these S-wave states are necessary since they are helpful to better understand their excited states both theoretically and experimentally. 

Accordingly, investigation of their weak, electromagnetic and strong decays are of much importance.
Specially, many theoretical and experimental efforts are concentrated on the radiative weak decays of heavy baryons, since they provide a possibility to investigate probes into the new physics beyond the standard model. Namely, LHC has produced a large number of heavy baryons and hyperons \cite{Cerri:2018ypt,Aaij:2017ddf,Junior:2018odx} and for the first time the rare radiative decay of $\Lambda^0_b\to\Lambda^0\gamma$ is observed by LHCb with a branching ratio of $(7.1\pm1.5\pm0.6\pm0.7)\times10^{-6}$ \cite{Aaij:2019hhx}. Furthermore, to explain the experimental data, there are some theoretical difficulties lasting for decades  \cite{Lach:1995we,Donoghue:1985rk}.

One of the most important classes of weak radiative decays is that based on $b \to s\gamma$ transition at quark level,  which is a flavor-changing neutral current (FCNC) process. In this process, based on the SM, the $W^{-}$ boson couples only to the left-handed quarks. Therefore the right-handed photons can be produced just by the helicity flips, meaning that the left- and right-handed amplitudes ratio is of order  $\mathcal{O}(m_s/m_b)$. Consequently, to investigate the presence of right-handed contributions, one has to measure branching fractions, angular and charge-parity-violating observables in $b \to s\gamma$ transitions. 
In this regard, the photon polarization can be investigated in the radiative decays of $b$-baryons. It is because there is no flavor mixing, the ground state spin is $1/2$ and also there are two spectator quarks. Therefore, these $b$-baryon decays alongside many studies on $B$-meson decays \cite{LHCb:2014vnw,Belle:2014sac,LHCb:2016oeh,LHCb:2019vks,LHCb:2020dof} can explore this field well.

After the observation of $\Lambda_{b}^{0}\to \Lambda\gamma$ decay \cite{Aaij:2019hhx}, recently, LHCb studied the $\Xi_{b}^{-}\to \Xi^{-}\gamma$ decay, which is also mediated by $b\to s\gamma$, and set an upper limit on its branching ratio that is ${\cal B}(\Xi_{b}^{-}\to \Xi^{-}\gamma) < 1.3 \times 10^{-4}$ \cite{LHCb:2021hfz}.
It is consistent with the one calculated using SU(3) flavor symmetry method, ${\cal B}(\Xi_{b}^{-}\to \Xi^{-}\gamma) = (1.23 \pm 0.64) \times 10^{-5}$ \cite{Wang:2020wxn}, but is in slight tension with the prediction of light-cone sum rules (LCSR) method, ${\cal B}(\Xi_{b}^{-}\to \Xi^{-}\gamma) = (3.03 \pm 0.10) \times 10^{-4}$ \cite{Liu:2011ema}.

In this paper we calculate the decay widths of the radiative decays of $\Xi_b^{0(-)}(\Xi^{'-}_{b})$ to $\Xi^{0(-)}$ and $\Sigma^{0(-)}$ baryons (i.e.,  $ \Xi_{b}^{-}\rightarrow\Xi^{-}\gamma  $, $ \Xi_{b}^{0}\rightarrow\Xi^{0}\gamma $, $ \Xi_{b}^{-}\rightarrow\Sigma^{-}\gamma  $, $ \Xi_{b}^{0}\rightarrow\Sigma^{0}\gamma $, $ \Xi_{b}^{'-}\rightarrow\Xi^{-}\gamma $ and $ \Xi_{b}^{'-}\rightarrow\Sigma^{-}\gamma $ transitions) via LCSR, and for the ones that their mean lifetimes are available, we also present the corresponding branching ratios. We show that the branching ratio for the $\Xi_{b}^{-}\to \Xi^{-}\gamma$ decay is consistent with the upper limit obtained by LHCb Collaboration \cite{LHCb:2021hfz} and the so-called tension with the previous LCSR prediction \cite{Liu:2011ema} is removed. In the analyses, form factors (FFs) are the main inputs of the problem and we use their values calculated via LCSR in full theory \cite{Azizi:2011mw}.

The organization of the paper is as follows. In next section,  we present the formalism  to calculate the corresponding decay widths and branching ratios in terms of FFs.  In  section 3 the numerical results are obtained. The last section is devoted to summary and conclusion.

\section{Formalism}\label{LH}

To investigate the weak radiative decays of $B_{b} \to B\gamma$, where $B_{b}$ is either of the heavy baryons $\Xi_b^{0(-)}$ or $\Xi^{'-}_{b}$, and $B$ is one of the outgoing $\Xi^{0(-)}$ or $\Sigma^{0(-)}$ baryons, one has to start with writing a suitable effective Hamiltonian defining these processes.
In this section, we present the effective Hamiltonian in terms of the relevant Wilson coefficients in the SM. Then, we define the transition amplitude and show how it depends on the FFs. Finally we write the decay width in terms of the corresponding FFs.

\subsection{\textit{The Effective Hamiltonian}}

The effective Hamiltonian in the SM and at the quark level for $b \to s\gamma$ and $b \to s g$ transitions can be written in terms of Wilson coefficients and local operators as \cite{Buchalla:1995vs}

\bea \label{Heff}
{\cal H}^{eff}&=& -\frac{G_{F}}{\sqrt{2}}V_{tb}V_{ts}^\ast 
\bigg[ {\sum\limits_{i=1}^{6}} C_{i}({\mu}) Q_{i}({\mu})+C_{7
	\gamma}(\mu) Q_{7\gamma}(\mu)+C_{8G}(\mu) Q_{8G}(\mu)\bigg].
\eea
Here $G_F$ is the Fermi coupling constant, $V_{ij}$ are the elements of the CKM matrix and $\mu$ is the QCD renormalization scale. The Wilson coefficients, $C_i$ and $C_{7\gamma,8g}$, may contain non-perturbative effects and can be viewed as the effective coupling constants whereas the local operators, $Q_i$ and $Q_{7\gamma,8g}$, can be considered as the effective vertices. The long-distance strong interaction contributions are coded in the local operators  and are given as
\bea
Q_{1}~\es ({\bar{s}}_{\alpha}c_{\beta} )_{V-A}
({\bar{c}}_{\beta} b_{\alpha})_{V-A},~\nnb \\
Q_{2}~\es({\bar{s}}_{\alpha}c_{\alpha})_{V-A}
({\bar{c}}_{\beta} b_{\beta} )_{V-A},~\nnb \\
Q_{3}~\es({\bar{s}}_{\alpha}b_{\alpha})_{V-A}\sum\limits_{q}
({\bar{q}}_{\beta} q_{\beta} )_{V-A},~\nnb \\
Q_{4}~\es({\bar{s}}_{\beta}b_{\alpha})_{V-A}\sum\limits_{q}
({\bar{q}}_{\alpha} q_{\beta} )_{V-A},~\nnb \\
Q_{5}~\es({\bar{s}}_{\alpha}b_{\alpha})_{V-A}\sum\limits_{q}
({\bar{q}}_{\beta} q_{\beta} )_{V+A},~\nnb \\
Q_{6}~\es({\bar{s}}_{\beta}b_{\alpha})_{V-A}\sum\limits_{q}
({\bar{q}}_{\alpha} q_{\beta} )_{V+A},~\nnb \\
Q_{7 \gamma}\es{e \over 4\pi^2} \bar{s}_{\alpha}\sigma^{\mu \nu}(m_b R+m_sL)b_{\alpha}\, F_{\mu \nu},~\nnb \\
Q_{8 G}\es {g_s \over 4\pi^2} \bar{s}_{\alpha}\sigma^{\mu \nu}(m_b R+m_sL)T^a_{\alpha \beta}b_{\beta}\,
G^{a}_{\mu \nu },~ \eea
where  $\alpha$  and $\beta$ are the color indices, and the right- and left-handed projectors are $R=(1+\gamma_5)/2$ and $L=(1-\gamma_5)/2$, respectively. In the last two operators, $e$ and $g_s$ are the electromagnetic and strong coupling constants, respectively.
Each operator corresponds to a certain interaction. $Q_{1,2}$ are current-current or tree operators, $Q_{3,4,5,6}$ are QCD penguin and $Q_{7\gamma,8g}$ the magnetic penguin operators. The electromagnetic field strength tensor $F_{\mu\nu}$ is defined as
\bea \label{emtensor}
 F_{\mu\nu}(x) &=& -i(\varepsilon_{\mu}q_{\nu}-\varepsilon_{\nu}q_{\mu})e^{iqx}~,
  \eea
where $\varepsilon_{\mu}$ and $q_{\nu}$ are the polarization 4-vector and momentum, respectively. For the $b \rightarrow s \gamma$ transition, the most relevant contribution comes from the magnetic penguin operator $Q_{7 \gamma}$ which reduces the effective Hamiltonian to 
\bea \label{Heff1} {\cal H}^{eff}(b \rightarrow s \gamma) &=& -{G_F e \over 4\pi^2\sqrt{2}} V_{tb}
V_{ts}^\ast  C^{eff}_{7}(\mu)\bar{s} \sigma_{\mu\nu} \Big[m_{b}R+m_{s}L \Big]bF^{\mu\nu}~. \eea
The relevant Wilson coefficient is $C^{eff}_{7}$ and, in the SM, it is given by \cite{Buras:1993xp}
\bea
\label{wilson-C7eff} C_7^{eff}(\mu_b) \es
\eta^{\frac{16}{23}} C_7(\mu_W)+ \frac{8}{3} \left(
\eta^{\frac{14}{23}} -\eta^{\frac{16}{23}} \right) C_8(\mu_W)+C_2 (\mu_W) \sum_{i=1}^8 h_i \eta^{a_i}~. \nnb\\ \eea
Here $\eta$ is defined as
 \bea \eta \es
\frac{\alpha_s(\mu_W)} {\alpha_s(\mu_b)}~,\eea
where 
\bea
\alpha_s(x)=\frac{\alpha_s(m_Z)}{1-\beta_0\frac{\alpha_s(m_Z)}{2\pi}\ln(\frac{m_Z}{x})},\eea
and $\alpha_s(m_Z)=0.118$ and $\beta_0=\frac{23}{3}$.
The coefficients $a_i$ and $h_i$ in Eq.\eqref{wilson-C7eff} have the following values:
\be\frac{}{}
\label{coefficients}
\begin{array}{rrrrrrrrrl}
	a_i = (\!\! & \f{14}{23}, & \f{16}{23}, & \f{6}{23}, & -
	\f{12}{23}, &
	0.4086, & -0.4230, & -0.8994, & 0.1456 & \!\!)  \vspace{0.1cm},\\
	h_i = (\!\! & 2.2996, & - 1.0880, & - \f{3}{7}, & - \f{1}{14}, &
	-0.6494, & -0.0380, & -0.0186, & -0.0057 & \!\!), 
\end{array}
\ee
and the coefficients  $C_{2,7,8}(\mu_W)$ are defined as
\bea
C_2(\mu_W)=1~,~~ C_7(\mu_W)=-\frac{1}{2}
D^\prime_0(x_t)~,~~ C_8(\mu_W)=-\frac{1}{2}
E^\prime_0(x_t)~ , \eea 
where 
\bea \label{Dprime0} D^\prime_0(x_t) \es -
\frac{(8 x_t^3+5 x_t^2-7 x_t)}{12 (1-x_t)^3}
+ \frac{x_t^2(2-3 x_t)}{2(1-x_t)^4}\ln x_t~, \\ \nnb \\
\label{Eprime0} E^\prime_0(x_t) \es - \frac{x_t(x_t^2-5 x_t-2)}
{4(1-x_t)^3} + \frac{3 x_t^2}{2 (1-x_t)^4}\ln x_t~. \eea

\subsection{\textit{Transition Amplitude}}

As an example, for the transition $\Xi_{b} \to \Xi\gamma$, by sandwiching the effective Hamiltonian between the initial heavy baryon $\Xi_{b}$ and the final baryon $\Xi$ states, one can obtain the corresponding amplitude as
\bea \label{Amplitude} {\cal M}^{(\Xi_{b}
	\rightarrow \Xi \gamma)}&=& \langle \Xi(p_{\Xi},s)|{\cal H}^{eff}(b \rightarrow s \gamma)|\Xi_{b}(p_{\Xi_{b}},s')\rangle ~, \eea
where $p_{\Xi}$ and $p_{\Xi_{b}}=p_{\Xi}+q$ are momenta of the $\Xi$ and $\Xi_{b}$ baryons; and $ s $ and $ s' $ are their spins, respectively.
To proceed, one has to write the transition amplitude in terms of relevant FFs. Inserting ${\cal H}^{eff}$ from Eq. \eqref{Heff1} to the amplitude \eqref{Amplitude}, one finds the corresponding matrix elements that can be written in terms of $f_2^{T}(0)$ and $g_2^{T}(0)$ FFs as follows:
\begin{eqnarray} \label{Transition-matrix-element}
\langle \Xi(p_{\Xi},s)|\bar s\;\sigma_{\mu\nu}q^\nu ( g_V &+& \gamma_5g_A ) b|\Xi_{b}(p_{\Xi_{b}},s')\rangle \nnb\\&=& \bar
u_{\Xi}(p_{\Xi},s)\sigma_{\mu\nu}q^\nu \Big(g_V f_2^{T}(0)+\gamma_5 g_A
g_2^{T}(0) \Big)u_{\Xi_{b}}(p_{\Xi_{b}},s').
\end{eqnarray}
Here $\bar u_{\Xi}(p_{\Xi},s)$ and $u_{\Xi_{b}}(p_{\Xi_{b}},s')$ are spinors of the $\Xi$ and $\Xi_{b}$ baryons, respectively; and $g_V=1+m_{s}/m_{b}$ and $g_A=1-m_{s}/m_{b}$.
The values of FFs will be taken from \cite{Azizi:2011mw}, which are calculated systematically for $ B_{b,c}\rightarrow B l^+l^- $ processes via LCSR in the full theory. The matrix elements defining the $ B_{b,c}\rightarrow B l^+l^- $ FCNC processes are defined in terms of twelve FFs in full QCD. Among them,  $f_2^{T}$ and $g_2^{T}$  are relevant to the radiative $B_{b} \to B\gamma$ transition.
In order to make this issue clear let us note that the FCNC $ \Xi_{b} \to \Xi l^+l^- $ transition, which is relevant to our present study, takes place via $ b\rightarrow s l^+l^- $ at quark level, whose effective Hamiltonian is given as
\begin{eqnarray} \label{ham} {\cal H}^{ b\rightarrow s l^+l^-}_{eff} \es \frac{G_F~\alpha_{em}
V_{tb}~V_{ts}^{^{*}}}{2\sqrt2~\pi} \Bigg\{\vphantom{\int_0^{x_2}}
C_{9}^{eff}~ \bar{s} \gamma_\mu (1-\gamma_5) b \bar l \gamma^\mu
 l +C_{10} ~\bar{s}
\gamma_\mu (1-\gamma_5) b \bar l \gamma^\mu \gamma_{5}l \nnb \\
\ek2 m_{b}~C_{7}^{eff}\frac{1}{q^{2}} ~\bar{s} i \sigma_{\mu\nu}q^{\nu}
(1+\gamma_5) b \bar l \gamma^\mu l \Bigg\}.
\end{eqnarray}
In order to get the relevant amplitude, we need to sandwich the effective Hamiltonian between the initial ($ \Xi_{b} $) and final ($ \Xi $) baryonic states. The latter effective Hamiltonian, contains two kinds of transition currents, namely $J_\mu^{tr,I}=\bar s \gamma_\mu
(1-\gamma_5) b$ and $ J_\mu^{tr,II}=\bar s i
\sigma_{\mu\nu}q^{\nu} (1+ \gamma_5) b$. Hence, the relevant matrix elements  are parameterized in terms of twelve form factors as
\begin{eqnarray}\label{matrixel1a} 
&&\langle \Xi(p_{\Xi},s) \mid  J_\mu^{tr,I} \mid  \Xi_{b}(p_{\Xi_{b}},s')   \rangle = \bar
{u}_{\Xi}(p_{\Xi},s) \Big[\gamma_{\mu}f_{1}(q^{2})+{i}
\sigma_{\mu\nu}q^{\nu}f_{2}(q^{2})
+ q^{\mu}f_{3}(q^{2})\nnb \\&-& \gamma_{\mu}\gamma_5
g_{1}(q^{2})-{i}\sigma_{\mu\nu}\gamma_5q^{\nu}g_{2}(q^{2}) -
q^{\mu}\gamma_5 g_{3}(q^{2}) \vphantom{\int_0^{x_2}}\Big] u_{ \Xi_{b}}(p_{\Xi_{b}},s')~,
\end{eqnarray}
and
\begin{eqnarray}\label{matrixel1b}
 &&\langle \Xi(p_{\Xi},s)\mid J_\mu^{tr,II} \mid  \Xi_{b}(p_{\Xi_{b}},s') \rangle
=\bar{u}_{\Xi}(p_{\Xi},s)
\Big[\gamma_{\mu}f_{1}^{T}(q^{2})+{i}\sigma_{\mu\nu}q^{\nu}f_{2}^{T}(q^{2})
+ q^{\mu}f_{3}^{T}(q^{2})\nnb \\ &+& \gamma_{\mu}\gamma_5
g_{1}^{T}(q^{2})+{i}\sigma_{\mu\nu}\gamma_5q^{\nu}g_{2}^{T}(q^{2}) +
q^{\mu}\gamma_5 g_{3}^{T}(q^{2}) \vphantom{\int_0^{x_2}}\Big]
u_{ \Xi_{b}}(p_{\Xi_{b}},s')~,
\end{eqnarray}
where  $f_i(q^{2})$, $g_i(q^{2})$,   $f^T_i(q^{2})$ and $g^T_i(q^{2})$ ($ i $ runs from $ 1 $ to $ 3 $) are twelve transition
form factors that are calculated via LCSR  in full QCD in Ref. \cite{Azizi:2011mw}.  For self-consistency of the paper, in Appendix, we briefly present how these form factors are calculated using LCSR.  Comparing the effective Hamiltonian of $ b\rightarrow s \gamma $ in Eq. (\ref{Heff1}) with the one in Eq. (\ref{ham}) for the $ b\rightarrow s l^+l^- $ transition, we see that  the operator $  \bar s i
\sigma_{\mu\nu}q^{\nu} (1+ \gamma_5) b$ and the Lorentz structures $ {i}\sigma_{\mu\nu}q^{\nu} $ and $ {i}\sigma_{\mu\nu}\gamma_5q^{\nu} $ on the right hand side of Eq. (\ref{matrixel1b}), and as a result, the form factors $ f_{2}^{T}(q^{2}) $  and $g_{2}^{T}(q^{2})  $ are relevant to the case of $ b\rightarrow s \gamma $. For real photon the values of form factors at  $ q^2=0 $ are used.

\subsection{\textit{Decay Width}}
	
Using the above mentioned matrix elements, one can calculate the total decay width, for $\Xi_{b} \to \Xi\gamma$ channel as an example, in terms of two FFs as 
\bea \label{DiffDecayRate}
\Gamma_{(\Xi_b\rightarrow\Xi\gamma)}=\frac{G_F^2 \alpha_{em}|V_{tb}V_{ts}^*|^2 m_b^2}{64\pi^4}|C_7^{eff}|^2 \left(\frac{m_{\Xi_b}^2-m_\Xi^2}{m_{\Xi_b}} \right)^3\Big(g_V^2
|f_2^{T}(0)|^2 + g_A^2 |g_2^{T}(0)|^2\Big),\nnb\\ \eea	
where $\alpha_{em}$ is the fine structure constant  evaluated at the $ Z $ mass scale. The FFs $f_2^{T}$ and $g_2^{T}$ are evaluated at $q^2=0$ since the outgoing photon is on-shell.
To obtain the corresponding branching ratio, one hast to multiply the total decay width by the lifetime of the initial heavy baryon $\Xi_b$ and divide by $\hbar$. Therefore it is possible to calculate the branching ratio of decays in which the lifetime of the initial heavy baryon is available.
	
\section{Numerical Results}\label{NA}

To perform the numerical analysis, we need some input parameters like the quark and baryon masses, some  physical constants, elements of CKM matrix and baryon lifetimes. They are collected in   Tables \ref{tab:mass} and \ref{tab:const}. As we previously mentioned, the main inputs of the analyses are the FFs, $f_2^{T}(0)$ and $g_2^{T}(0)$. We use their values from Ref. \cite{Azizi:2011mw} obtained using LCSR in full theory  and presented in Table \ref{tab:FFs}. 
\begin{table}[ht]
	\centering
	\rowcolors{1}{lightgray}{white}
	\begin{tabular}{cc}
		\hline \hline
		Particle   &  mass    
		\\
		\hline \hline
		$ m_{s} $            &     $ 93^{+11}_{-5}   $ $\text{MeV}$      \\
		$ m_{b} $            &    $ (4.78 \pm0.06)    $ $\text{GeV}$     \\
		$ m_{t} $            &    $ (172.76 \pm 0.30) $ $\text{GeV}$     \\
		\hline
		$ m_{W} $            &    $ (80.379 \pm 0.012)$ $\text{GeV}$     \\
		\hline
		$ m_{\Xi_{b}^{-}} $            &    $ (5797.0 \pm 0.6)$ $\text{MeV}$     \\
		$ m_{\Xi_{b}^{0}} $            &    $ (5791.9 \pm 0.5)$ $\text{MeV}$     \\
		$ m_{\Xi_{b}^{'-}} $            &    $ (5935.02 \pm 0.05)$ $\text{MeV}$     \\
		\hline
		$ m_{\Xi^{-}} $            &    $ (1321.71 \pm 0.07)$ $\text{MeV}$     \\
			$ m_{\Xi^{0}} $            &    $ (1314.86 \pm 0.20)$ $\text{MeV}$     \\
			$ m_{\Sigma^{-}} $            &    $ (1197.45 \pm 0.04)$ $\text{MeV}$     \\
			$ m_{\Sigma^{0}} $            &    $ (1192.642 \pm 0.024)$ $\text{MeV}$     \\
		\hline \hline
	\end{tabular}
	\caption{The values of quark, $ W $ boson and baryon masses \cite{PDG}.}\label{tab:mass}
\end{table}	
\begin{table}[ht]
	\centering
	\rowcolors{1}{lightgray}{white}
	\begin{tabular}{cc}
		\hline \hline
		Constant   &  Value    
		\\
		\hline \hline
		$ \hbar $            &     $ 6.582\times10^{-22}   $ $\text{MeV~s}$      \\
		$ G_F $            &     $ 1.166\times10^{-5}   $ $\text{GeV}^{-2}$      \\
		$ \alpha_{em} $            &     $ 1/137  $       \\
		$|V_{ts}| $            &     $ (38.8\pm 1.1)\times10^{-3}$       \\
        $|V_{tb}| $            &     $ 1.013\pm 0.030$      
        \\
        $\tau_{\Xi_b^{-}} $            &     $ (1.572 \pm 0.040)\times 10^{-12}$ $\text{s}$      
        \\
        $\tau_{\Xi_b^{0}} $            &     $ (1.480 \pm 0.030)\times 10^{-12}$ $\text{s}$      
        \\
		\hline \hline
\end{tabular}
\caption{The values of physical constants, elements of CKM matrix and baryon lifetimes \cite{PDG}.}\label{tab:const}
\end{table}
\begin{table}[ht]
	\centering
	\rowcolors{1}{lightgray}{white}
	\begin{tabular}{ccc}
		\hline \hline
		Decay  &  $f_2^{T}(0)$  &  $g_2^{T}(0)$   
		\\
		\hline \hline
		$ \Xi_b \to \Xi\gamma $    &   $0.157\pm 0.041$  &  $0.155 \pm 0.040$    \\
		$ \Xi_b \to \Sigma\gamma $    &   $0.049 \pm 0.012$  &  $0.013 \pm 0.003$    \\
		$ \Xi^{'}_b \to \Xi\gamma $    &   $0.109 \pm 0.028$  &  $0.024 \pm 0.006$    \\
		$ \Xi^{'}_b \to \Sigma\gamma $    &   $0.138 \pm 0.036$  &  $0.085 \pm 0.021$    \\

	\hline \hline
\end{tabular}
\caption{The values of corresponding FFs at $q^2=0$ \cite{Azizi:2011mw}.}\label{tab:FFs}
\end{table}
As we also previously noted, in Ref. \cite{Azizi:2011mw}, all the twelve form factors defining the $ B_{b}\rightarrow B l^+l^-  $ transitions are calculated as  functions of $ q^2 $ in full theory and using the most general forms of the interpolating currents for the initial and final baryonic states. Among them, the two form factors  $f_2^{T}(q^2)$ and  $g_2^{T}(q^2)$ at $ q^2=0 $ are needed in the numerical analyses of the radiative $ B_{b}\rightarrow B\gamma $ channels under consideration, which are presented in Table \ref{tab:FFs}. The form factors $f_2^{T}(q^2)$ and  $g_2^{T}(q^2)$ only at $ q^2=0 $ are also calculated in Ref. \cite{Liu:2011ema} using specific currents for the participating baryons that contain only one possible diquark and the third quark attached, for each case. In Ref. \cite{Liu:2011ema}, the variations of form factors at $ q^2=0 $ with respect to the Borel parameter are presented, but  their final numerical values with related uncertainties are not given explicitly. Hence, we can not include them in  Table \ref{tab:FFs} in order to compare their values with the values taken from Ref. \cite{Azizi:2011mw}.

Inserting the numerical values of the parameters, one can find the total decay width of the corresponding decays. For the decays that the lifetime of the initial heavy baryon is available (those with initial states $\Xi_b^{0(-)}$), one can easily find the branching ratios. The values of total decay widths and branching ratios as well as the existing results for  these parameters  in SU(3) flavor symmetry method and previous LCSR (Ref. \cite{Liu:2011ema}) are presented in Tables \ref{tab:width} and  \ref{tab:Br}. In Table \ref{tab:Br}, the upper limit for the branching ratio of the radiative $ \Xi_{b}^{-}\rightarrow\Xi^{-}\gamma  $ mode recently set by LHCb Collaboration is also presented. From Tables \ref{tab:width} and \ref{tab:Br}, we see that our prediction on the width and branching ratio of $ \Xi_{b}\rightarrow\Xi\gamma  $ channel  differs considerably from the previous LCSR result provided by Ref. \cite{Liu:2011ema}. This difference can be attributed to different interpolating currents, different input parameters as well as different working windows for the auxiliary parameters that  Refs. \cite{Liu:2011ema} and  \cite{Azizi:2011mw} use in the analyses. As it is clear from Tables \ref{tab:width} and \ref{tab:Br}, we calculate the width and branching ratio of more radiative modes with initial $ \Xi_b $ and $ \Xi^{'}_b $ baryons that were not considered by  Ref. \cite{Liu:2011ema}.
\begin{table}[ht]
	\centering
	\rowcolors{1}{lightgray}{white}
	\begin{tabular}{ccc}
		\hline \hline
		Decay  &  $\Gamma[\text{GeV}] $ (Present Work)  &  $\Gamma[\text{GeV}] $ (LCSR)  \cite{Liu:2011ema}
		\\
		\hline \hline
        $ \Xi_{b}^{-}\rightarrow\Xi^{-}\gamma  $ & $4.52^{+2.66}_{-2.05}\times 10^{-18}$ & $ (1.34\pm0.07) \times 10^{-16} $ 
        \\
        $ \Xi_{b}^{0}\rightarrow\Xi^{0}\gamma $ & $4.52^{+2.65}_{-2.04}\times10^{-18}$ & -
        \\
        $ \Xi_{b}^{-}\rightarrow\Sigma^{-}\gamma  $ & $ 2.57^{+1.41}_{-1.10}\times10^{-19} $ & -
        \\
        	$ \Xi_{b}^{0}\rightarrow\Sigma^{0}\gamma $  & $ 2.57^{+1.40}_{-1.10}\times10^{-19} $ &  -
        \\
        $ \Xi_{b}^{'-}\rightarrow\Xi^{-}\gamma $ & $ 1.31^{+0.76}_{-0.59}\times 10^{-18} $ & - 
        \\
        $ \Xi_{b}^{'-}\rightarrow\Sigma^{-}\gamma $ & $ 2.78^{+1.61}_{-1.25}\times 10^{-18} $ & - 
        \\
		\hline \hline
\end{tabular}
\caption{Decay widths for different channels obtained in the present study compared to the one from previous LCSR.}\label{tab:width}
\end{table}
\begin{table}[ht]
	\centering
	\rowcolors{1}{lightgray}{white}
	\begin{tabular}{ccccc}
		\hline \hline
		Decay   &  ${\cal B}$ (Present Work)  & ${\cal B}$ (LCSR)  \cite{Liu:2011ema}  & ${\cal B}$ (SU(3))\cite{Wang:2020wxn} & ${\cal B}$ (Experiment)\cite{LHCb:2021hfz}
		\\
		\hline \hline
        $ \Xi_{b}^{-}\rightarrow\Xi^{-}\gamma  $  & $ 1.08^{+0.63}_{-0.49} \times 10^{-5} $ &$ (3.03 \pm 0.10) \times 10^{-4} $& $(1.23\pm0.64)\times10^{-5}$ & $ < 1.3 \times 10^{-4} $
        \\
        $ \Xi_{b}^{0}\rightarrow\Xi^{0}\gamma $ & $1.02^{+0.60}_{-0.46} \times 10^{-5} $ & -&$(1.16\pm0.60)\times10^{-5}$ & -
        \\
        $ \Xi_{b}^{-}\rightarrow\Sigma^{-}\gamma  $  & $ 6.14^{+3.36}_{-2.63} \times 10^{-7}$ &- &$(5.74\pm3.21)\times10^{-7}$ & -
        \\
        	$ \Xi_{b}^{0}\rightarrow\Sigma^{0}\gamma $   &  $ 5.77^{+3.16}_{-2.47} \times 10^{-7}  $ &-& $(2.71\pm1.50)\times10^{-7}$ & -
        \\
		\hline \hline
\end{tabular}
\caption{Branching ratios for different channels obtained in the present study compared with the results of previous LCSR,  SU(3) flavor symmetry method  and existing experimental data.}\label{tab:Br}
\end{table}

It can be easily  seen from Table \ref{tab:Br} that our result on the branching ratio of $\Xi_{b}^{-}\rightarrow\Xi^{-}\gamma$ decay lies below the upper limit put by  LHCb and there is no any tension. We  also see  good agreements among our results and those  of SU(3) flavor symmetry method for this channel and other presented modes. Our results for the branching ratios of the presented channels as well as the the decay widths of the channels with initial $ \Xi_{b}^{'-}$ baryon, whose lifetime is not available from the experiment, may shed light on future related experiments.

\section{Summary and Conclusions}\label{SC}

Recently, the LHCb Collaboration put an upper limit on the branching ratio of the $ \Xi_{b}^{-}\rightarrow\Xi^{-}\gamma $ decay: ${\cal B}(\Xi_{b}^{-}\to \Xi^{-}\gamma) < 1.3 \times 10^{-4}$ \cite{LHCb:2021hfz}. Although it is consistent with prediction of  the SU(3) flavor symmetry method, ${\cal B}(\Xi_{b}^{-}\to \Xi^{-}\gamma) = (1.23 \pm 0.64) \times 10^{-5}$ \cite{Wang:2020wxn}, it  is in a tension with the previous prediction made by the method of LCSR, ${\cal B}(\Xi_{b}^{-}\to \Xi^{-}\gamma) = (3.03 \pm 0.10) \times 10^{-4}$ \cite{Liu:2011ema}. Inspired by this, we investigated this decay mode and other   radiative decays of  $\Xi_b^{0(-)}(\Xi^{'-}_{b})$ to $\Xi^{0(-)}$ and $\Sigma^{0(-)}$ baryons ($ \Xi_{b}^{0}\rightarrow\Xi^{0}\gamma $, $ \Xi_{b}^{-}\rightarrow\Sigma^{-}\gamma  $, $ \Xi_{b}^{0}\rightarrow\Sigma^{0}\gamma $, $ \Xi_{b}^{'-}\rightarrow\Xi^{-}\gamma $ and $ \Xi_{b}^{'-}\rightarrow\Sigma^{-}\gamma $) using the FFs calculated in  light-cone QCD sum rules in full theory. The values of FFs  were borrowed from Ref. \cite{Azizi:2011mw}, which are  systematically obtained for $ B_{b,c}\rightarrow B l^+l^- $ processes via LCSR in full QCD without any approximation. The matrix elements defining the $ B_{b,c}\rightarrow B l^+l^- $ FCNC processes are defined in terms of twelve FFs in full theory. Two of these FFs, namely  $f_2^{T}(q^2)$ and $g_2^{T}(q^2)$ for real photon at $q^2=0$, are involved in each radiative process considered in the present study.

We obtained the decay width for all of the considered channels as well as branching ratio for those that the lifetime of the initial state is experimentally available. Our result on  ${\cal B}(\Xi_{b}^{-}\to \Xi^{-}\gamma) = 1.08^{+0.63}_{-0.49} \times 10^{-5}$, lies below the upper limit set by LHCb and we witness no tension. For this channel and  some other modes, our results agree well with the predictions of  the SU(3) flavor symmetry method, as well. The difference between our result on branching ratio of $ \Xi_{b}^{-}\rightarrow\Xi^{-}\gamma $ with prediction of Ref. \cite{Liu:2011ema} may be attributed to the fact that the two studies use different interpolating currents, different input parameters as well as different working windows for the auxiliary parameters in the analyses.

Our results on widths and branching ratios presented in Tables \ref{tab:width} and \ref{tab:Br} can shed light on future experiments. They may also be  checked via other phenomenological approaches.

\section{Appendix: LCSR for form factors}	

In this appendix, we give some details of calculations for the form factors $f^T_i(q^{2})$ and $g^T_i(q^{2})$, as an example for   $ \Xi_b^0\rightarrow\Xi^0 $ channel and by  using the tensor transition current $  J_\mu^{tr,II}=\bar s i
\sigma_{\mu\nu}q^{\nu} (1+ \gamma_5) b$. The starting point is to consider the following correlation function: 
\begin{eqnarray}\label{T}
\Pi^{II}_{\mu}(p,q) = i\int d^{4}xe^{iqx}\langle \Xi^0(p) \mid
T\{ J_\mu^{tr,II}(x), \bar J^{\Xi_b^0}(0)\}\mid
0\rangle~,
\end{eqnarray}
 where  $J^{\Xi^0_b}$ is the
interpolating current for  the initial 
$\Xi^0_b$  baryon (hereafter we omit the electric charge) and $ T $ is the time-ordering operator. Considering all quantum numbers, the interpolating current in its general form is given as
\begin{eqnarray}\label{IPF}
 J^{\Xi_b} &=& {1\over \sqrt{6}} \epsilon^{abc} \Big\{
2 \Big( u^{aT} C s^b \Big) \gamma_5 b^c + 2 \beta \Big( u^{aT}
C \gamma_5 s^b \Big) b^c + \Big( u^{aT} C b^b \Big) \gamma_5
s^c + \beta \Big(u^{aT} C
\gamma_5 b^b \Big) s^c \nnb \\
&+& \Big(b^{aT} C s^b \Big) \gamma_5 u^c + \beta \Big(b^{aT} C
\gamma_5 s^b \Big) u^c \Big\}~,
\end{eqnarray}
where $C$ is the charge conjugation operator;  $a$,
$b$ and $c$ are the color indices
and $\beta$ is an
arbitrary mixing parameter, which is fixed based on the standard prescriptions of the LCSR method.   The value $\beta=-1$ corresponds to the Ioffe
current.

We calculate this correlation function once in terms of hadronic parameters, then, in terms of QCD degrees of freedom in deep Euclidean region and the final baryon distribution amplitudes (DAs). In hadronic window, the above mentioned correlation function is saturated by a complete set of $ \Xi_b$ baryon. By performing the four integral over $ x $, we get
\begin{eqnarray}
\label{phys1111}
 \Pi_{\mu}^{II}(p,q)&=&\sum_{s}\frac{\langle
\Xi(p)\mid J^{tr,II}_\mu
\mid \Xi_b(p+q,s')\rangle\langle \Xi_b(p+q,s')\mid
\bar J^{\Xi_b}(0) \mid
0\rangle}{m_{\Xi_b}^{2}-(p+q)^{2}} +\cdots~,
\end{eqnarray}
where  $...$ stands for the contributions of the higher states
and continuum. To proceed, besides the transition matrix elements defined in terms of the corresponding form factors in the body text, we need to introduce the residue $\lambda_{\Xi_b}$ through the matrix element,
\begin{eqnarray}\label{matrixel2} \langle \Xi_b(p+q,s')\mid
\bar J^{\Xi_b}(0) \mid
0\rangle=\lambda_{\Xi_b}
\bar u_{\Xi_b}(p+q,s')~.
\end{eqnarray}
Using all the  definitions in Eq.(\ref{phys1111}) requires applying the  completeness relation for spin--1/2 Dirac particle as
\begin{eqnarray}\label{spinor}
\sum_{s'}u_{\Xi_b}(p+q,s')\overline{u}_{\Xi_b}(p+q,s')=
\not\!p+\not\!q+m_{\Xi_b}~,
\end{eqnarray}
which leads to  final representation of the correlation function in the  hadronic side  in terms of the  related form factors and other hadronic parameters:
\begin{eqnarray}\label{sigmaaftera}  \Pi_{\mu}^{II}(p,q)\es
\frac{\lambda_{\Xi_b}u_{\Xi}(p,s)}{m_{\Xi_b}^{2}-(p+q)^{2}}
\Big\{ 2 f_{1}^{T}(q^{2}) p_\mu + 2 f_{2}^{T}(q^{2})p_\mu\not\!q+
\Big[ f_2^{T}(q^2) + f_3^{T}(q^2)\Big]q_\mu\not\!q \nnb \\
\ek 2g_1^{T}(q^2)p_{\mu}\gamma_5 - 2
g_2^{T}(q^2)p_\mu\not\!q\gamma_5
-\Big[g_2^{T}(q^2)+g_3^{T}(q^2) \Big]q_\mu\not\!q\gamma_5 \nnb \\
\ar \mbox{\rm other structures} \Big\}+... ~,
\end{eqnarray}
To find the form factors, one needs to select the corresponding Lorentz structures, coefficients of which will be equated to the ones in the QCD side of the correlation function.

On QCD side, we insert the explicit form of the current $J^{\Xi_b}$ into the correlation function Eq. (\ref{T}), after contracting the quark fields using Wick's theorem, we get
\begin{eqnarray}\label{mut.mm}
\Pi^{II}_{\mu} &=& \frac{-i}{\sqrt{6}} \epsilon^{abc}\int d^4x
e^{-iqx} \Bigg\{\Bigg(\Big[2( C )_{\phi\eta}
(\gamma_5)_{\rho\beta}+( C )_{\phi\beta}
(\gamma_5)_{\rho\eta}+(C)_{\beta\eta}(\gamma_5)_{\rho\phi}\Big]
+\beta\Bigg[2(C \gamma_5 )_{\phi\eta}(I)_{\rho\beta}
 \nonumber \\
&+& (C \gamma_5 )_{\phi\beta}(I)_{\rho\eta}+(C \gamma_5
)_{\beta\eta}(I)_{\rho\phi} \Bigg]\Bigg) \Big[
i\sigma_{\mu\nu}q^\nu(1-\gamma_5)
\Big]_{\sigma\theta}\Bigg\}S_b(-x)_{\beta\sigma}\langle 0 |
s_\eta^a(0) s_\theta^b(x)
u_\phi^c(0) | \Xi (p)\rangle,\nonumber\\
\end{eqnarray} 
where, $ S_b(x) $  is the heavy quark propagator given by
\begin{eqnarray}\label{heavylightguy}
 S_b (x)& =&  \frac{m_{b}^{2}}{4\pi^{2}}\frac{K_{1}(m_{b}\sqrt{-x^2})}{\sqrt{-x^2}}-i
\frac{m_{b}^{2}\not\!x}{4\pi^{2}x^2}K_{2}(m_{b}\sqrt{-x^2})\nnb \\ &- &i g_s \int \frac{d^4 k}{(2\pi)^4}
e^{-ikx} \int_0^1 dv \Bigg[\frac{\not\!k + m_b}{( m_b^2-k^2)^2}
G^{\mu\nu}(vx) \sigma_{\mu\nu} 
+ \frac{1}{m_b^2-k^2} v x_\mu G^{\mu\nu} \gamma_\nu \Bigg],\nnb \\
\end{eqnarray}
with $K_i$ being the Bessel functions of the second kind. In Eq. (\ref{mut.mm}), $ \langle 0 |
s_\eta^a(0) s_\theta^b(x)
u_\phi^c(0) | \Xi (p)\rangle $ is the wave function of $ \Xi $ baryon. It is parameterized in terms of different calligraphic functions ($ \mathcal{S}_i $,$ \mathcal{P}_i $, $ \mathcal{V}_i $, $ \mathcal{A}_i $ and $ \mathcal{T}_i $), mass and momentum of the corresponding baryon and Dirac matrices and is given generally  for spin-1/2 octet $ B $ baryons as 
\begin{eqnarray}\label{wave func}
&&4\langle0|\epsilon^{abc}{q_1}_\alpha^a(a_1 x){q_2}_\beta^b(a_2
x){q_3}_\gamma^c(a_3 x)|B(p)\rangle
=\mathcal{S}_1m_{B}C_{\alpha\beta}(\gamma_5B)_{\gamma}+
\mathcal{S}_2m_{B}^2C_{\alpha\beta}(\rlap/x\gamma_5B)_{\gamma}\nnb\\
\ar \mathcal{P}_1m_{B}(\gamma_5C)_{\alpha\beta}B_{\gamma}+
\mathcal{P}_2m_{B}^2(\gamma_5C)_{\alpha\beta}(\rlap/xB)_{\gamma}+
(\mathcal{V}_1+\frac{x^2m_{B}^2}{4}\mathcal{V}_1^M)(\rlap/pC)_{\alpha\beta}(\gamma_5B)_{\gamma}
\nnb\\\ar
\mathcal{V}_2m_{B}(\rlap/pC)_{\alpha\beta}(\rlap/x\gamma_5B)_{\gamma}+
\mathcal{V}_3m_{B}(\gamma_\mu
C)_{\alpha\beta}(\gamma^\mu\gamma_5B)_{\gamma}+
\mathcal{V}_4m_{B}^2(\rlap/xC)_{\alpha\beta}(\gamma_5B)_{\gamma}\nnb\\\ar
\mathcal{V}_5m_{B}^2(\gamma_\mu
C)_{\alpha\beta}(i\sigma^{\mu\nu}x_\nu\gamma_5B)_{\gamma} +
\mathcal{V}_6m_{B}^3(\rlap/xC)_{\alpha\beta}(\rlap/x\gamma_5B)_{\gamma}
+(\mathcal{A}_1\nnb\\
\ar\frac{x^2m_{B}^2}{4}\mathcal{A}_1^M)(\rlap/p\gamma_5
C)_{\alpha\beta}B_{\gamma}+
\mathcal{A}_2m_{B}(\rlap/p\gamma_5C)_{\alpha\beta}(\rlap/xB)_{\gamma}+
\mathcal{A}_3m_{B}(\gamma_\mu\gamma_5 C)_{\alpha\beta}(\gamma^\mu
B)_{\gamma}\nnb\\\ar
\mathcal{A}_4m_{B}^2(\rlap/x\gamma_5C)_{\alpha\beta}B_{\gamma}+
\mathcal{A}_5m_{B}^2(\gamma_\mu\gamma_5
C)_{\alpha\beta}(i\sigma^{\mu\nu}x_\nu B)_{\gamma}+
\mathcal{A}_6m_{B}^3(\rlap/x\gamma_5C)_{\alpha\beta}(\rlap/x
B)_{\gamma}\nnb\\\ar(\mathcal{T}_1+\frac{x^2m_{B}^2}{4}\mathcal{T}_1^M)(p^\nu
i\sigma_{\mu\nu}C)_{\alpha\beta}(\gamma^\mu\gamma_5
B)_{\gamma}+\mathcal{T}_2m_{B}(x^\mu p^\nu
i\sigma_{\mu\nu}C)_{\alpha\beta}(\gamma_5 B)_{\gamma}\nnb\\\ar
\mathcal{T}_3m_{B}(\sigma_{\mu\nu}C)_{\alpha\beta}(\sigma^{\mu\nu}\gamma_5
B)_{\gamma}+
\mathcal{T}_4m_{B}(p^\nu\sigma_{\mu\nu}C)_{\alpha\beta}(\sigma^{\mu\rho}x_\rho\gamma_5
B)_{\gamma}\nnb\\\ar \mathcal{T}_5m_{B}^2(x^\nu
i\sigma_{\mu\nu}C)_{\alpha\beta}(\gamma^\mu\gamma_5 B)_{\gamma}+
\mathcal{T}_6m_{B}^2(x^\mu p^\nu
i\sigma_{\mu\nu}C)_{\alpha\beta}(\rlap/x\gamma_5
B)_{\gamma}\nnb\\
\ar
\mathcal{T}_7m_{B}^2(\sigma_{\mu\nu}C)_{\alpha\beta}(\sigma^{\mu\nu}\rlap/x\gamma_5
B)_{\gamma}+
\mathcal{T}_8m_{B}^3(x^\nu\sigma_{\mu\nu}C)_{\alpha\beta}(\sigma^{\mu\rho}x_\rho\gamma_5
B)_{\gamma}~, 
\end{eqnarray}
 where $ B $ denotes the spinor of the related baryon in the right hand side of the above expresion. The calligraphic functions in Eq. (\ref{wave func}) have not definite twists, but, they can be written in terms of the $ B $ baryon DAs of definite twists. The expressions of these calligraphic functions   and their relations with the scalar, pseudoscalar, vector, axial vector and tensor DAs for $ \Xi $ baryon together with all the related parameters and constants are given in  Refs. \cite{Azizi:2011mw,Liu:2009uc}. 
 
 Inserting the heavy quark propagator and the wave function of the $ \Xi $ baryon  into Eq. (\ref{mut.mm}) and performing the lengthy but straightforward calculations, one finds the expression of the correlation function on QCD side in momentum space. Matching the coefficients of the corresponding Lorentz structures both from the hadronic and QCD sides of the correlation function leads to the expressions for the corresponding form factors. To suppress the contributions of the higher states and continuum and enhance the ground state contribution, we apply the Borel transformation and continuum subtraction according to the standard prescriptions of the LCSR method. These procedures brings two  more auxiliary parameters, namely, the Borel mass parameter and continuum threshold that together with the previously introduced mixing parameter $ \beta $ in the initial baryon's current are fixed in  Ref. \cite{Azizi:2011mw} based on the standard criteria like the dominance of the ground state over the higher states and continuum and convergence of the light cone series obtained. The working intervals of these helping parameters are used to find the $ q^2 $ dependence of the form factors. As we mentioned in the body text, the values of the form factors $ f_2^T (q^2=0) $ and $ g_2^T (q^2=0) $ are needed in this study that are given in Table \ref{tab:FFs} for the decay modes under consideration.
 
\section*{Acknowledgment}
	K. Azizi is thankful to Iran Science Elites Federation (Saramadan)
for the partial  financial support provided under the grant number ISEF/M/400150.

\end{document}